\documentclass[fleqn,12pt,twoside]{article}
\usepackage[headings]{espcrc1}

\readRCS
$Id: espcrc1.tex,v 1.2 2004/02/24 11:22:11 spepping Exp $
\usepackage{graphicx}
\usepackage[figuresright]{rotating}

\newcommand{\AmS}{{\protect\the\textfont2
  A\kern-.1667em\lower.5ex\hbox{M}\kern-.125emS}}

\title{\large\bf Elasticity of nuclear medium as a principal macrodynamical promoter of electric pygmy dipole  resonance}

\author{S.I. Bastrukov\address[MCSD]{Laboratory of Informational Technologies,\\
        Joint Institute for Nuclear Research, 141980 Dubna, Russia}%
        \thanks{Institute of Astronomy,
                National Tsing Hua University, 30013 Hsinchu, Taiwan.},
        I.V. Molodtsova\addressmark{},
        \c S. Mi\c sicu\address{National Institute for Nuclear Physics, Bucharest, P. O. Box MG6, Romania},
         H.-K. Chang\address{Department of Physics and Institute of Astronomy,\\ National Tsing Hua University, 30013 Hsinchu, Taiwan}
             and
         D.V. Podgainy\addressmark[MCSD]}

\runtitle{Preprint of paper accepted for publication in Physics Letters B}
\runauthor{S.I. Bastrukov, I.V. Molodtsova, \c S. Mi\c sicu, H.-K. Chang, D.V. Podgainy}

\begin{document}

% typeset front matter
\maketitle

\begin{abstract}
Motivated by arguments of the nuclear core-layer model formulated in [S.I. Bastrukov, J.A. Maruhn, Z.Phys. A 335 (1990) 139], the macroscopic excitation mechanism of the electric pygmy dipole resonance (PDR) is considered as owing its origin to perturbation-induced effective decomposition of a nucleus into two spherical domains - undisturbed inner region treated as a static core and dynamical layer undergoing elastic shear vibrations.
 The elastic restoring force is central to the excitation mechanism under consideration and has the same physical meaning as in macroscopic model of nuclear giant resonances involving distortions of the Fermi-sphere providing unified description of isoscalar giant electric and magnetic resonances of multipole degree $\ell\geq 2$ in terms of two fundamental vibrational modes in an elastic sphere, to wit,
 as spheroidal (electric) and torsional (magnetic) modes of shear elastic oscillations of the nodeless field of material displacements excited in the entire nucleus volume. In the present paper focus is placed on
 the emergence of dipole overtone in the frequency spectrum of spheroidal
 elastic vibrations as Goldstone soft mode. To emphasis this feature of dipole resonant excitation
 imprinted in the core-layer model we
  regain spectral equation for the frequency of spheroidal elastic vibrations trapped in the finite-depth layer,
  derived in the above paper, but using canonical equation of an elastic continuous medium. The obtained analytic equations for the frequency of dipole vibrational state in question
  and its excitation strength lead to the following estimates for the PDR energy centroid  $E_{PDR}(E1)=[31\pm 1]\, A^{-1/3}$ MeV and the total excitation probability
  $B_{PDR}(E1)=[1.85\pm 0.05]\,10^{-3}\, Z^2\, A^{-2/3}$ $e^2$ fm$^2$
  throughout the nuclear chart exhibiting fundamental character of this soft dipole mode of nuclear resonant response.
\end{abstract}



 \section{Introduction}
 It is generally recognized today that macroscopic behavior of a nucleus at excitation
 of giant resonances of multipole degree $\ell\geq 2$ located lower than the compressional giant monopole and the
 giant dipole resonances bears strong resemblance to elastic shear (non-compressional) oscillations of a solid globe. Such an understanding, prompted long ago by work of Bertsch \cite{B-74}, has come into existence during the past three decades as a result of numerous investigations clearly indicating that macroscopic description of
 giant-resonant nuclear response in terms of shear oscillations of an elastic sphere provides proper account of experimentally observed size effect -- smooth variation of integral parameters of isoscalar giant resonances throughout the nuclear chart, such as centroid of energy, spread width and total excitation strength (e.g. \cite{IJMP-A-07} and references therein). This feature is generally thought of as exhibiting fundamental character of nuclear giant-resonant response \cite{HVDW-01}. In this latter context, a great deal of current interest centers on the electric pigmy dipole resonance (PDR) which is observed by the Nuclear Resonance Fluorescence (NRF) technique as a concentration of electric dipole strength near the neutron threshold (e.g. [4-16] and references therein), that is, in the energy domain where nuclear resonance-like excitations exhibit features generic to shear oscillations of an elastic sphere.

 It is the subject of the present paper to investigate elastodynamic excitation mechanism of
 the electric PDR, that is, as owing its origin to elasticity of nuclear medium.
 In so doing we focus on physical arguments and technical details expounded in Ref.\cite{BM-90}
 in which the effect of perturbation-induced effective decomposition of nucleus
 into two spherical domains - undisturbed by perturbation inner region, treated as a static
 core, and peripheral dynamical layer set in non-rotational elastic oscillations
 has been studied within the framework of macroscopic model of giant resonances involving distortions of the Fermi-sphere (e.g.[18-25]). One of the prime purposes of work \cite{BM-90} was to formulate mathematically trackable procedure of evaluating a fractional part of the nucleus volume involved in
 the elastic shear vibrations detected as giant isoscalar $E\ell$ resonances.
 Emphasis was laid on the core-layer model
 predictions regarding the energies of quadrupole and octupole overtones of shear elastic oscillations.
 By varying the depth of layer, which serves as an input parameter of the model, it was found that the deeper layer the higher is the excitation energy.
 In this paper we focus on dipole overtone of the layer-against-core elastic oscillations
 and accentuate a macroscopic mechanism of emergence of dipole vibrational
 excitation imprinted in the core-layer model as Goldstone soft vibrational mode. The most conspicuous feature
 of such a mode is that it can exist then and only then
 when elastic oscillations turn out trapped in the peripheral layer of finite depth, not in the entire volume of the nucleus.

\section{Basic equations}

 In the model under consideration a nucleus is thought of as
 an ultra fine spherical piece of elastic continuous medium condensed to the normal nuclear density
 $\rho$ and uniform distribution of the electric charge density $\rho_e$.
 The nucleus response to an external perturbation (induced by inelastically scattered electrons or elastically scattered gamma-quanta in NFR measurements) is described by the field of material displacements $u_i$ which serves as a basic dynamical variable of intrinsic collective fluctuations of nucleons.
 For non-compression fluctuations,  $\delta \rho=-\rho\nabla_k u_k=0$,
 the equation describing elastic dynamics of nuclear material in the nucleus volume reads
\begin{eqnarray}
  \label{e2.1}
  &&\rho {\ddot u}_i=\nabla_k\, \sigma_{ik},\quad \sigma_{ik}=2\mu\,u_{ik},\quad  u_{ik}=\frac{1}{2}(\nabla_i
  u_k+\nabla_k  u_i),\quad u_{kk}=\nabla_k\,u_k=0
  \end{eqnarray}
 where constant $\mu$ is the shear modulus relating applied shear stress $\sigma_{ik}$ to resulting shear stain $u_{ik}$, the Hooke's law of elastic deformation. The conservation of energy is described by equation
 \begin{eqnarray}
  \label{e2.2}
  \frac{\partial }{\partial t}\int \frac{\rho {\dot u}^2}{2}\,d{\cal V}=
  -2\int \mu\, u_{ik}{\dot u}_{ik}d{\cal V}.
   \end{eqnarray}
The fluctuating fields of material displacement $u_i$ and shear deformations $u_{ik}$ can be conveniently represented
in the following separable form
\begin{eqnarray}
  \label{e2.3}
  u_i({\bf r},t)=a_i({\bf r})\alpha(t),\quad u_{ik}=a_{ik}\alpha(t),\quad a_{ik}=\frac{1}{2}[\nabla_i a_k+\nabla_k a_i]
  \end{eqnarray}
 where $a_i({\bf r})$ stands for the time-independent field of instantaneous displacements and amplitude $\alpha(t)$
 defines time evolution of intrinsic elastic distortions. Following the line of argument of Refs.\cite{IJMP-A-07} and \cite{BM-90}, we focus on perturbation-induced shear fluctuations of the nucleus material in which the field of displacement obeys the vector Laplace equation
 \begin{eqnarray}
  \label{e2.4}
  \nabla^2 {\bf u}({\bf r},t)=0,\quad\quad \nabla\cdot{\bf u}({\bf
  r},t)=0
  \end{eqnarray}
 which is characteristic equation of the quasi-static regime of elastic oscillations thought of as
 long wavelength limit of standing-wave regime governed by the Helmholtz equation,
 $\nabla^2 {\bf u}+k^2{\bf u}=0$, because in the limit of long wavelength $\lambda\to \infty$ (and, hence, $k=2\pi/\lambda\to 0$) this latter equation is reduced to (\ref{e2.4}) \cite{IJMP-A-07}.
 Two fundamental solutions of (\ref{e2.4}), given by the even parity poloidal
 and odd parity toroidal  solenoidal fields of fundamental basis \cite{CHANDRA}, built on the general solution to the scalar Laplace equation $\nabla^2 \chi({\bf r})=0$ laid the ground for generally accepted Lamb's
 classification of vibrational eigenstates  in an elastic sphere. Namely,
 the even-parity spheroidal and odd-parity torsional vibrational states.
 In what follows we confine our discussion to spheroidal vibrational mode in which the poloidal
 field of instantaneous material displacement is irrotational and can be represented, therefore,
 as follows \cite{IJMP-A-07}:
 \begin{eqnarray}
 \label{e2.7}
 && {\bf a}_s({\bf r})=\nabla \,\chi({\bf r}\,\quad\quad \chi({\bf r})=f_\ell({\bf r})P_\ell(\cos\theta),\quad\quad
 f_\ell({\bf r})=[{\cal A}_\ell\,r^\ell+{\cal
  B}_\ell\,r^{-(\ell+1)}].
 \end{eqnarray}
 The structure of this field is identical to that utilized in hydrodynamical model assuming irrotational flow of
 incompressible fluid that was in the past a fairly successful idea adopted to describe other modes of very collective and energetic nuclear excitations.
 Also, it is noteworthy that the radial function $f_\ell(r)$ has no nodes, from what the term nodeless oscillations is derived. The arbitrary constants ${\cal A}_\ell$ and ${\cal B}_\ell$ are eliminated from boundary conditions motivated by physical arguments.

 \begin{figure}
 \centering{\includegraphics[width=12.cm]{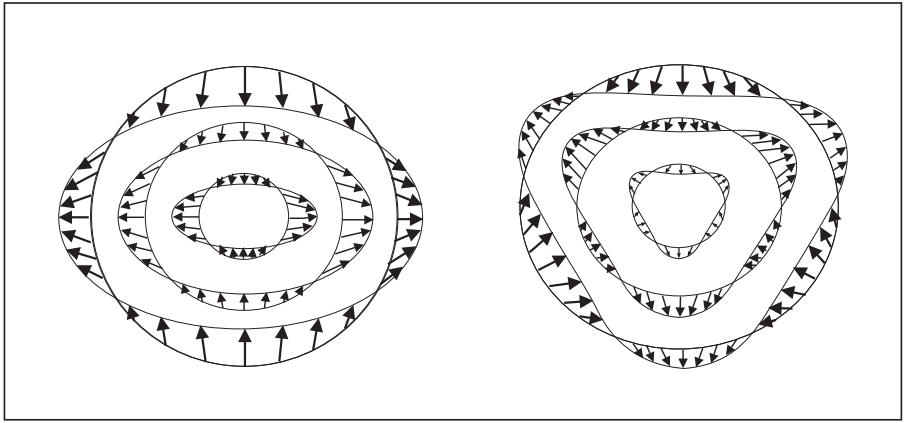}}
 \caption{\footnotesize{The irrotational fields of material displacements picturing distribution of elastic stresses
 whose emergence is attributed to resistivity of circular orbits of single-particle Fermi-motion in the mean field of shell model to perturbation-induced distortions of their equilibrium shapes. The picture
 visualizes quadrupole (left) and octupole (right) overtones of nodeless elastic oscillations
 excited in the entire nucleus volume and detected as isoscalar $E2$ and $E3$ giant resonances.}}
\end{figure}

 In Fig.1, the incessant Fermi-motion of independent nuclear quasi-particles
 in the mean field potential of shell model is visualized by regularly ordered circular orbits which serve
 as basic explanatory devices of the macroscopic model under consideration. The fast process of nuclear resonant excitations is associated with release of
 short-time electromagnetic load resulting in quasi-static oscillations of orbits about their equilibrium shapes under the action of restoring force of elastic shear stress, as pictured in this figure for quadrupole and octupole
 nodeless spheroidal elastic shear vibrations.

\begin{figure}
\centering{\includegraphics[width=12.cm]{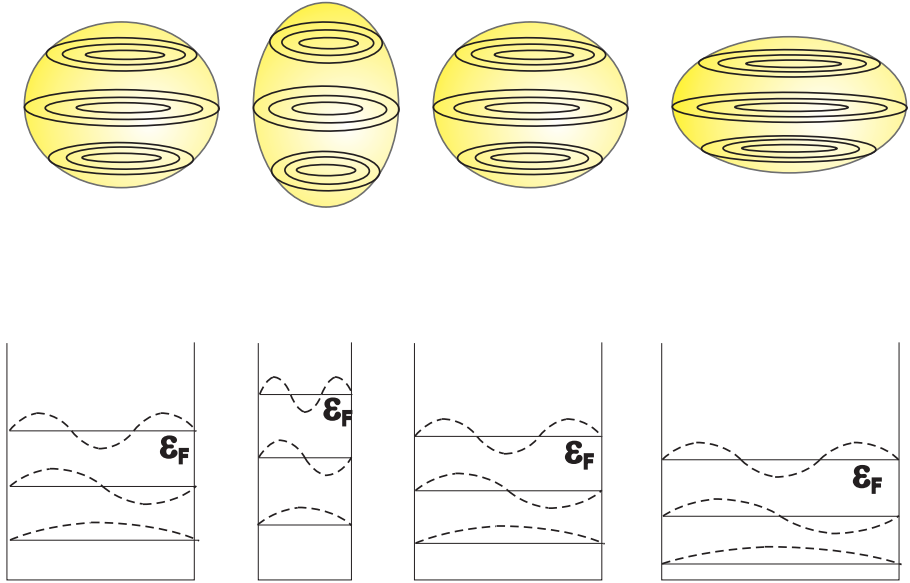}}
 \caption{\footnotesize{Anisotropic distortions of single-particle orbitals forming the nuclear Fermi-sphere in momentum space and corresponding squeeze-spread distortions of the nuclear mean field potential
 in the fast process of non-compressional
 elastic vibrational response of nucleus detected as isoscalar giant resonances.
 The most conspicuous feature of this response is that generic to the ground state the shell's order in filling of the mean field potential by single-particles states and
 original nodal structure of wave functions of single-particle states is preserved.}}
\end{figure}

 Fig.2 illustrates theoretical treatment of
 quasi-static regime of nodeless elastic shear vibrations from the viewpoint of the distorted Fermi-sphere model
 and canonical shell-model picture. The most prominent feature of non-compressional resonant nuclear response
 is that the original Fermi-distribution of nucleon quasi-particles in the momentum space and original shell-ordered sequence of discrete states of
 Fermi-motion of independent quasi-particles in the potential of nuclear mean field are not affected.

 The frequency $\omega$ of quasi-static regime of nodeless spheroidal oscillations is
 uniquely computed by the Rayleigh's energy variational method (see, for details, \cite{IJMP-A-07}) which leads to the equation for temporal amplitude $\alpha(t)$ having the form of equation of harmonic oscillations
   \begin{eqnarray}
\label{e2.9}
 && \frac{d{\cal E}}{dt}=0,\quad
 {\cal E}=\frac{M{\dot\alpha}^2}{2}+\frac{K\alpha^2}{2},\quad\to\quad
 {\ddot \alpha}(t)+\omega^2\alpha(t)=0,\quad \omega^2=\frac{K}{M}
\end{eqnarray}
 with the inertia $M$ and stiffness $K$ given by
 \begin{eqnarray}
  \label{e2.10}
  && M=\int \rho\, a_i({\bf r})\,a_i({\bf r}) d{\cal V},\quad\quad
  K=2\int \mu\, a_{ik}({\bf r})\,a_{ik}({\bf r}) d{\cal V}.
 \end{eqnarray}
 The experimentally measured energy centroid of isoscalar electric resonance $E(E\ell)$ is
 identified with the energy of $\ell$-pole spheroidal oscillations with frequency $\omega_s(\ell)$;
 the link between these two quantities is given by standard quantum-mechanical equation
 \begin{eqnarray}
 \label{e2.12}
 E(E\ell)=\hbar\omega_s(\ell),\quad\quad \omega_s(\ell)=\sqrt\frac{K(\ell)}{M(\ell)}.
 \end{eqnarray}
 Bearing in mind that non-compressional oscillations of an ultra fine electrically charged spherical mass
 of nuclear medium are accompanied by oscillations of the charge-current density $\delta {\bf j}=\rho_e\,{\dot {\bf u}}_s$ (where ${\dot {\bf u}}_s$ stands for the rate of displacements in spheroidal mode of nodeless elastic oscillations) and that the integral characteristics of corresponding vibrational states are the electric moments of charge-current density ${\cal M}_j(E\ell)$,  the electric excitation strength of the $\ell$-pole nuclear response to the long-wavelength electromagnetic field can be evaluated by standard equation of the macroscopic electrodynamics of continuous media
  \begin{eqnarray}
   \label{e2.14}
 B(E\ell)=(2\ell+1)<|{\cal M}_j(E\ell)|^2>,\quad\quad {\cal M}_j(E\ell)=\frac{i}{\omega (\ell+1)}\int \delta {\bf j}
 \cdot \nabla\, r^\ell P_{\ell}(\theta) \,d{\cal V}.
 \end{eqnarray}
 The last two formulae provide computational basis for physical interpretation of nuclear giant-resonant
 excitations in continuum-mechanical terms of nodeless shear vibrations of charged elastic sphere.

  \section{Nuclear response by spheroidal elastic oscillations trapped in the peripheral finite-depth layer}

 To evaluate a fractional part of the nucleus volume involved in
 elastic shear vibrations, detected as giant isoscalar $E\ell$ resonances, in \cite{BM-90}
 the nucleus response has been considered in the core-layer model
 presuming the perturbation-induced decomposition of nucleus into effective static core and dynamical layer undergoing nodeless shear oscillations which are controlled by elastic restoring force.
 The term effective means that two-component, core-layer, picture emerges solely in the process
 of excitation, not in the ground state, so that the very notion of core should be thought, thereby, of as
 reflecting the dynamically inert central region of nucleus unaffected by perturbation and remaining at rest.
 To get better understanding dynamical peculiarities of elastodynamic mechanism of giant-resonant excitations, an analytic calculation of spectral equation for the frequency
 has been performed in the approximation of sharp edge and homogeneous material parameters, to wit,
 the density $\rho$ and the shear modulus $\mu$ of nuclear medium.
 The depth of dynamical layer involved in elastic vibrations can be conveniently represented as $\Delta R=R(1-x)$, where $x=R_c/R$, is the basic parameter of the core-layer model regulating dependence of the energy and the excitation strength of vibrational state upon the layer depth.
 The obtained in \cite{BM-90} field of oscillating non-rotational
 material displacements in the peripheral nuclear layer can be represented as follows
\begin{eqnarray}
\label{e3.1}
 && {\bf u}_s({\bf r},t)={\bf a}_s({\bf r})\,\alpha(t)\quad\quad {\bf a}_s=\nabla \chi({\bf r})\quad \chi({\bf
 r})=[{\cal A}_\ell\,r^\ell+{\cal B}_\ell\,r^{-(\ell+1)}]P_\ell(\cos\theta),\\
 \label{e3.2}
 && {\cal A}_\ell=\frac{{\cal N}_\ell}{\ell(\ell+1)},\quad
  {\cal B}_\ell=-\frac{{\cal N}_\ell}{\ell(\ell+1)}\,R_c^{2\ell+1},\quad
  {\cal N}_\ell=\frac{R^{\ell+3}}{R^{2\ell+1}-R_c^{2\ell+1}}.
  \end{eqnarray}
 where constants ${\cal A}_\ell$ and ${\cal B}_\ell$ have been eliminated from boundary conditions of
 impenetrability of perturbation in the core, $u_r\vert_{r=R_c}=0$, and compatibility of the rate of
 displacements with the rate of
 spheroidal distortions of the nucleus surface:
 ${\dot u}\vert_{r=R}={\dot R}(t)$, where $R(t)=R[1+{\alpha}_\ell(t)P_\ell(\cos\theta)]$.
 The inertia $M_s(\ell,x)$ and the stiffens $K_s(\ell,x)$ computed as functions of multipole degree $\ell$
 and parameter $x$ are given by
  \begin{figure}
 \centering{\includegraphics[width=10.cm]{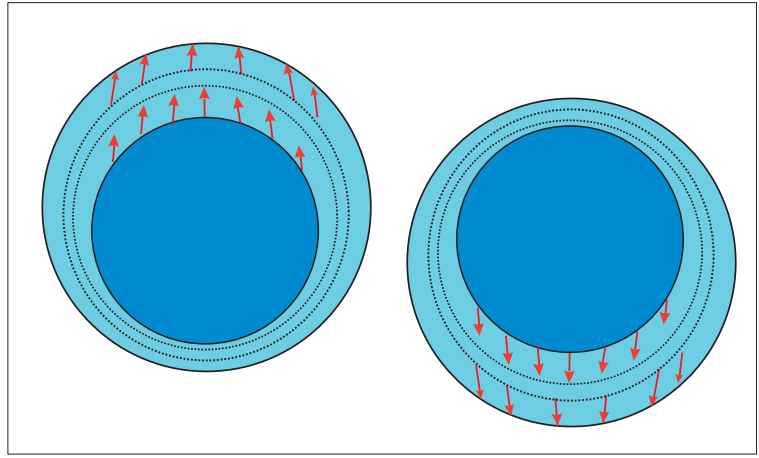}}
 \caption{\footnotesize{Artist view of nuclear elastic distortions in suggested macroscopic mechanism of isoscalar
 electric pygmy dipole resonance as elastic dipole soft mode. The excitation process
 is thought of as an effective decomposition of nucleus, induced by elastically scattered gamma-quanta of FNR
 technique,  into two domains  -- undisturbed by perturbation
 internal spherical region treated, thereby, as a static core and peripheral dynamic layer
 undergoing differentially translational oscillations driven by restoring force of elastic shear stresses. The
 emergence of elastic force is attributed to resistivity to disruption of peripheral circular periodic orbit of incessant Fermi-motion of independent quasi-particles in the nuclear mean field of shell model.}}
\end{figure}

  \begin{eqnarray}
  \label{e3.3}
  && M_s(\ell,x)
=\frac{4\pi R^5\rho}{\ell(2\ell+1)(1-x^{2\ell+1})}
\left[1+\frac{\ell}{(\ell+1)}x^{2\ell+1}
\right],\\
 \label{e3.4}
&& K_s(\ell,x) = \frac{8\pi R^3
\mu}{(1-x^{2\ell+1})^2}\left[
\frac{(\ell-1)}{\ell}(1-x^{2\ell-1}) +
\frac{(\ell+2)}{(\ell+1)}x^{2\ell-1}\,(1-x^{2\ell+3})\right].
 \end{eqnarray}
 The resultant frequency spectrum $\omega_s(\ell,x)$ reads
    \begin{eqnarray}
 \label{e3.5}
 &&\omega^2_s(\ell,x)=\omega_0^2\left\{\frac{2(2\ell+1)}{(1-x^{2\ell+1})}
 \left[\frac{(\ell^2-1)(1-x^{2\ell-1})+\ell(\ell+2)x^{2\ell-1}(1-x^{2\ell+3})}
 {(\ell+1)+\ell x^{2x+1}}\right]\right\}.
 \end{eqnarray}
 It follows when the core radius $R_c\to 0$ and, hence parameter, $x=(R_c/R)\to 0$, a limiting
 case when the entire volume of nucleus sets in oscillations, the last spectral
 formula takes the form
 \begin{eqnarray}
 \label{e3.6}
 \omega_s(\ell)=\omega_0[2(2\ell+1)(\ell-1)]^{1/2},\quad\quad \omega_0=\frac{c_t}{R}
 \end{eqnarray}
 showing that the lowest overtone is of quadrupole, $\ell=2$, degree;
 $c_t=[\mu/\rho]^{1/2}$ is the speed of transverse shear wave in the bulk nuclear matter.
 However, when $x\neq 0$ the lowest overtone, as is easily seen, is of dipole, $\ell=1$, degree.
 In this latter case a peripheral layer executes elastic differentially translational shear oscillations relative to static core, as pictured in Fig.3.

 \section{Dipole soft mode of elastic layer-against-core shear oscillations}

 It is appropriate for the former to discuss the obtained analytic formulae for the
 energy and excitation strength of dipole overtone of nodeless elastic layer-against-core oscillations
 by highlighting the emergence of dipole overtone as Goldstone soft mode.
 The mass parameter and stiffness of dipole differentially translational oscillations of the finite-depth
 layer against static core have the form
    \begin{eqnarray}
\label{e3.7}
 && M_s(\ell=1,x)=\frac{4\pi R^5\rho}{3}\frac{(1+x^3/2)}{(1-x^3)},\quad
  K_s(\ell=1,x)
=12\pi R^3 \mu  \frac{x(1-x^5)}{(1-x^3)^2}
 \end{eqnarray}
 and corresponding energy of dipole vibrational state is given by

  \begin{figure}
 \centering{\includegraphics[width=10.cm]{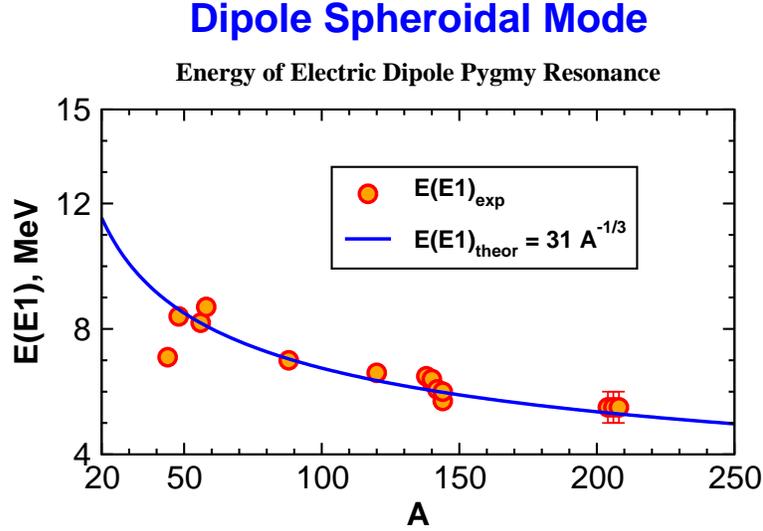}}
 \caption{\footnotesize{Theoretically computed energy of electric dipole soft mode of elastic translational oscillations of
 layer against core in juxtaposition with data on energy centroid of the low-energy E1 strength taken from [6-16].}}
 \end{figure}

\begin{eqnarray}
  \label{e3.8}
  && E(E1,x)=\hbar\omega_s(\ell=1,x)\quad \omega_s(\ell=1,x)=\omega_0\left[\frac{9\,x(1-x^5)}{(1-x^3)(1+x^3/2)}\right]^{1/2}.
 \end{eqnarray}
 One sees again that when, $x=0$, the coefficient of vibrational rigidity vanishes: $ K_s(\ell=1,x=0)=0$
 and, as follows from Hamiltonian of oscillator, the total absorbed energy goes in kinetic energy
 of the center-of-mass motion.  This simple argument shows that
 the dipole excitation in question can exist as vibrational mode when perturbation sets
 in differentially translational fluctuations solely
 peripheral nuclear layer of finite depth leaving the central spherical region of nucleus unaltered.
 Such behavior is typical for the Goldstone soft modes whose most conspicuous feature is that the mode disappears (the  frequency turns into zero), when one of parameters of vibrating system tends to zero: $\omega(\ell=1,x)\to 0$,
 when $x\to 0$.

  The total dipole strength of electromagnetic response computed as squared dipole moment of the
  charge-current density fluctuations excited in the surface finite-depth layer is given by

   \begin{figure}
 \centering{\includegraphics[width=10.cm]{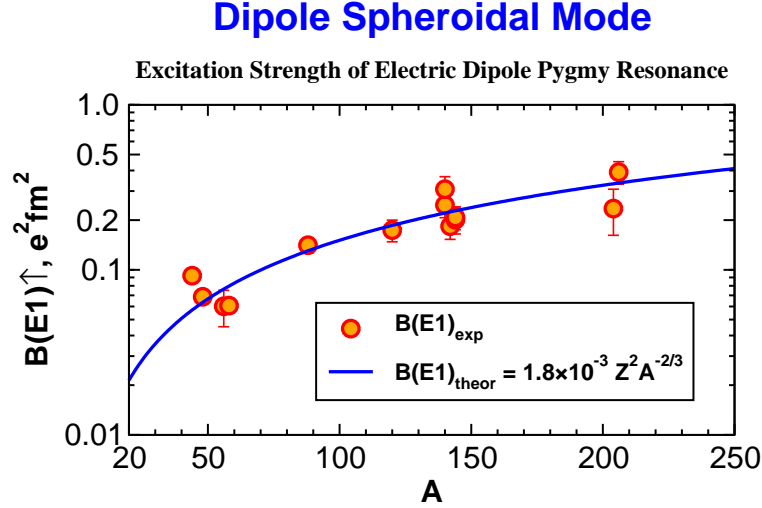}}
  \caption{\footnotesize{The strength of low-energy E1 electromagnetic nuclear resonant
  response. Symbols - data of works [6-16] and line -- is E1 excitation strength of the dipole electric soft mode
  of elastic oscillations computed as integral electric dipole moment of charge-density current
 in this vibration state.}}
\end{figure}

 \begin{eqnarray}
 \label{e3.9}
 && B(E1,x)=3<\vert {\cal M}_j(E1)\vert>=\frac{\tilde\rho_e^2}{\rho} \frac{\hbar\,R^3}{2\,\omega_0}    \left[\frac{(1-x^3)^3}{x(1-x^5)(1+x^3/2)}\right]^{1/2}.
 \end{eqnarray}
 By $\tilde\rho_e$ we denote the charge density of the peripheral layer which in the model
 of homogeneous layer can be defined as $\tilde\rho_e=\gamma\,\rho_e$
 where $\rho_e=(Z/A)en$ stands for the average charge density of nucleus as a whole with $n$ being the average
 particle density of nucleons. Physically, the parameter of fractional charge density, $\gamma < 1$,
 takes into account the neutron-dominated content of nuclear matter in the surface layer of nucleus.

 The considered macroscopic mechanism of emergence of dipole vibrational mode can be regarded as having
 universal character (generic to all stable nuclei of nuclear chart) if the input parameters of the model, namely, the speed of transverse shear wave $c_t$, geometrical parameter $x$, and fractional charge density, $\gamma$ would
 have one and the same values for all nuclei. If so, from the obtained equations for the energy centroid and excitation strength it follows that the integral characteristics smoothly vary with mass number $A$
 and this variation is given by typical for the giant resonances estimates:  $E(E1,x)=\kappa_E(x)\, A^{-1/3}$ and $B(E1,x)=\kappa_B(x)\,Z^2A^{-4/3}$, respectively, where $\kappa_E(x)$ and $\kappa_B(x)$ are constants and the
 link between atomic number $Z$ and mass number $A$ is given by the well-known empirical formula: $Z=A[2+0.015A^{2/3}]^{-1}$.
 Bearing this in mind (and that the elastodynamic excitation mechanism provide proper account of
 isoscalar giant resonances with $\ell\geq 2$, as discussed below) it is tempting to consider
 the available experimental data on the low-energy electric PDR in the context of above predictions of the core-layer model for the energy and excitation strength of the dipole soft mode.
 By varying parameter $x$ so as to attain best agreement with
 data on the energy centroid of PDR as a function of mass number, we get $x=0.33$.
 Having fixed this parameter and applying the obtained formula for the excitation strength to data on total excitation strength one finds that the fractional charge density is
 given by $\gamma=6.6\,10^{-2}$. The net outcome is summarized by the following estimates
  \begin{eqnarray}
 \label{e3.11}
 E_{PDR}(E1)=[31\pm 1]\, A^{-1/3}\, {\rm MeV},\quad B_{PDR}(E1)=[1.85\pm 0.05]\,10^{-3}\, Z^2\, A^{-2/3} e^2 {\rm fm}^2.
 \end{eqnarray}
 showing that the electric PDR is fundamental resonant mode of nuclear response generic to all stable nuclei of nuclear chart, as it is demonstrated in Fig.4 and Fig.5 where theoretically computed
 energy centroid and total excitation strength of elastic dipole soft vibrational mode are plotted in
 juxtaposition with experimental data for the electric PDR borrowed form [6-16].

 A special comment that deserves to be made is that the difference between proposed elastodynamic excitation mechanism of the electric PDR implying the isoscalar nature of this soft mode and the hydrodynamic mechanism lying at the base of isovector macroscopic model \cite{MDB-71,SIS-90} from the standpoint of which the electric
  PDR is thought of as low-frequency counterpart of giant dipole resonance. In calculations reported in \cite{MDB-71} the low-energy dipole resonant mode arises as a solution of fairly sophisticated equations describing oscillations of relative neutron-proton density driven by restoring force defined as gradient of symmetry energy and, thus, implying the isovector type and compressional character of material oscillations.
  In the meantime, numerous theoretical investigations of nuclear giant resonances by macroscopic methods of the
  theory of material continua, developed over the past three decades,  unambiguously indicate
  that excitation of compression nuclear vibrations demands much more energy of electromagnetic perturbation
  than that goes to excitation of the electric pygmy dipole resonance.

\section{Discussion and summary}

 The line of argument presented above shows that the electric PDR  emerges as a soft dipole mode of elastic shear
 oscillations of irrotational flow which turn out to be confined in the finite-depth surface layer.
 These two signatures are the main features distinguishing the low-energy electric PDR from the electric toroid-dipole resonant (TDR) mode centered at $E_{TDR}(E1)\sim 70\,A^{-1/3}$ MeV [27-33].
 The characteristic peculiarities of TDR mode  (considered in \cite{BMS-93} on the same physical footing,
 that is, as driven by restoring force of shear elastic distortions) is that in this latter case
 a nucleus responds by oscillations of rotational, i.e. vortical, flow with the torus-like shape of the flow lines
 and such oscillations are excited in the whole volume of nucleus.

\begin{figure}
 \centering{\includegraphics[width=12.cm]{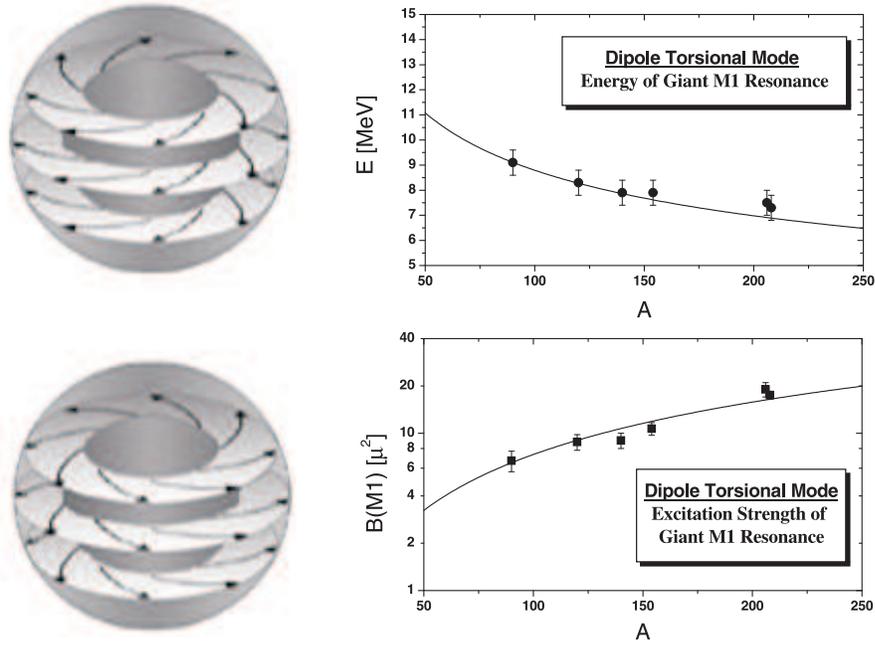}}
 \caption{\footnotesize{Artist view of nuclear elastic distortions in magnetic dipole resonant (MDR)
 response of nuclei which is described as
 caused by effective decomposition of nucleus, induced by elastically scattered gamma-quanta of FNR
 technique,  into two regions -- undisturbed by perturbation static core and peripheral dynamic layer
 undergoing torsional, differentially rotational, oscillations driven restoring force of elastic shear stresses. One sees that theoretical predictions of the core-layer model \cite{BMS-95,BLM-97} for the energy centroid,
 $E(MDR)=41\,A^{-1/3}$, MeV and excitation strength, $B(MDR)=8.5\, 10^{-2}\,Z^2\,A^{-2/3}\,\, \mu_N^2$
 of magnetic dipole resonance adequately reproduce experimental data \cite{RFH-91} throughout the nuclear chart.}}
\end{figure}

   It is noteworthy that based on arguments similar to expanded above, in Refs.\cite{BMS-95,BLM-97}
  it was shown that magnetic dipole resonance (MDR), experimentally detected by NRF technique as well \cite{RFH-91},
  can also be interpreted as a result of perturbation-induced core-layer decomposition of nucleus, but accompanied by
  differentially rotational, torsional, elastic oscillations of peripheral layer relative to static core, as pictured in Fig.6. To this end, it worth emphasizing that macroscopic elastodynamic treatment of low-frequency
  dipole nuclear resonant excitations provides a remarkable way of unifying understanding of the electric pygmy dipole resonance and magnetic dipole resonance as soft modes of differentially translational (PDR)
  and differentially rotational (MDR) elastic oscillations of the finite-depth layer against static core,
  respectively, the oscillations driven by one and the same restoring force of shear deformations.
  This point of view is strengthened by unified elastodynamic interpretation of isoscalar electric $E\ell$ and magnetic $M\ell$ giant resonances of multipole degree $\ell\geq 2$ in terms of two fundamental, spheroidal and
  torsional, vibrational modes in a viscoelastic sphere, excited in the entire nucleus volume: the electric giant resonances are treated as even parity spheroidal (shake) mode of nodeless elastic shear oscillations with frequency $\omega_s(\ell)$ of irrotational vector field of material displacements and
  the magnetic giant resonances as odd-parity torsional (twist) mode of nodeless
  shear oscillations with frequency  $\omega_t(\ell)$ of differentially rotational vector field of material
  displacements, respectively, as pictured in Fig.7.

  \begin{figure}[ht!]
 \centering{\includegraphics[width=16.cm]{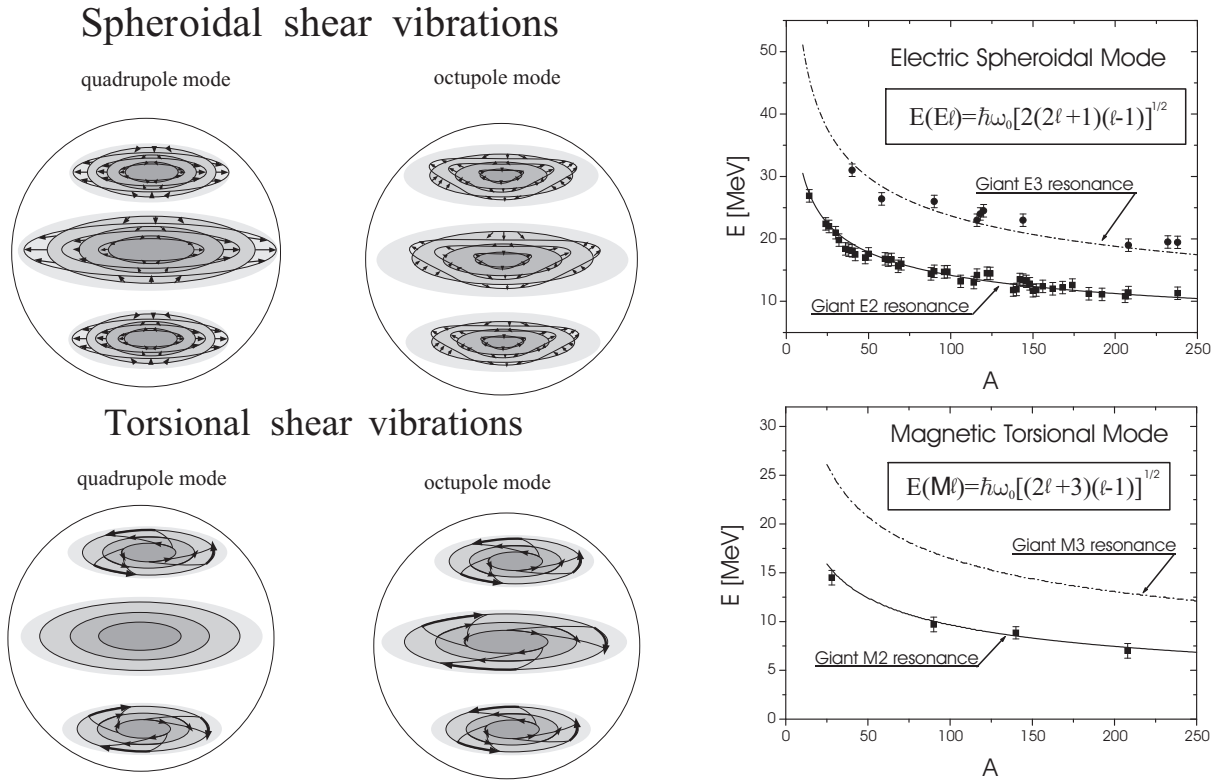}}
 \caption{\footnotesize{Intrinsic distortions in the spheroidal and torsional modes
  of global elastic shear vibrations of nucleus detected as giant electric and magnetic resonances.
  Energy centroid of electric $E\ell$ and magnetic $M\ell$ resonances of multipole
  degree $\ell\geq$ as a function of mass number $A$ computed with $\omega_0=c_t/R$ where $c_t=[\mu/\rho]^{1/2}$ is the speed of transverse wave of elastic shear and $R=r_0\,A^{1/3}$ is the nucleus radius.}}
\end{figure}

  The predictions of the nuclear solid-globe model regarding energies $E=\hbar\omega$ and spread widths $\Gamma=\hbar\tau^{-1}$ of electric $E\ell$ and magnetic $M\ell$ giant resonances of multipole degree
  $\ell\geq$ treated in terms of above spheroidal and torsional nodeless shear
  vibrations of a viscoelastic sphere are given by the following spectral equations \cite{IJMP-A-07}
   \begin{eqnarray}
 \label{e3.11}
  && E(E\ell)=\hbar\,\omega_0[2(2\ell+1)(\ell-1)]^{1/2},\quad\quad \Gamma
 (E\ell)=\frac{\hbar}{\tau_0}[(2\ell+1)(\ell-1)]^{-1},\\[0.2cm]
   \label{e3.12}
  && E(M\ell)=\hbar\,\omega_0[(2\ell+3)(\ell-1)]^{1/2},\quad\quad \Gamma (M\ell)=\frac{\hbar}{2\tau_0}[(2\ell+3)(\ell-1)]^{-1},\\
  \label{e3.14}
  && E\sim \omega_0=\frac{c_t}{R}=\frac{\sqrt{\mu/\rho}}{R}\sim A^{-1/3},\quad\quad \Gamma\sim \tau_0^{-1}=\frac{\eta}{\rho R^2}\sim A^{-2/3}.
 \end{eqnarray}
 where $\omega_0$ and $\tau_0$ are the natural units of frequency and lifetime of shear vibrations
 restored by bulk force of elastic stresses and damped by friction force of shear stresses, the quantities
 carrying information about shear modulus $\mu$ and shear viscosity $\eta$ of nuclear matter.
 The obtaining of such information has been and still is among the main purposes of macroscopic modeling of nuclear giant resonance in terms of vibrations of an ultra small mass of nuclear continuous medium. From identification of theoretically computed
 energy of spheroidal quadrupole vibrational state with
 experimental energy of giant $E2$ resonance, taken from \cite{B-81}, it follows $\mu\simeq 4\times 10^{33}$ dyn cm$^{-2}$. The fact that predictions of the model fairly accurately match the data on $E3$ (from \cite{TY-81}) and $M2$ (from \cite{LNCNR-96}) with no use of any adjustable constants, as one can see in Fig.7, lends strong support to elastodynamical macroscopic mechanism of excitation of isoscalar giant resonances.
 Also it seems appropriate to add that a similar line of argument has been utilized to extract the shear viscosity $\eta$ of nuclear matter from systematic data on spread width of giant resonances and
 the work of Hasse \cite{H-78} is among the first to point out that such procedure leads to $\eta\simeq 3\cdot 10^{11}$\,\, dyn sec cm$^{-2}$ (see, for detail, \cite{IJMP-A-07}).
 These latter inferences of nuclear physics regarding transport coefficients of shear elasticity $\mu$ and shear viscosity $\eta$ of nuclear material is of particular
 interest for current investigations on asteroseismology of neutron stars, an actively developing domain
 of pulsar astrophysics studying quake-induced seismic vibrations of neutron stars.
 Such vibrations are detected as quasi-periodic oscillations of electromagnetic emission from these cosmic
 nuclear matter objects exhibiting similar elastodynamical pattern of their vibrational behavior (e.g. [43-45] and references therein).

\section{Acknowledgements}

 This work is partly supported by NSC of Taiwan,
 under grants  NSC-96-2628-M-007-012-MY3 and NSC-97-2811-M-007-003.


\begin{thebibliography}{<num>}


\bibitem{B-74} G.F. Bertsch,  Ann. Phys. 86 (1974) 138

  G.F. Bertsch, R.A. Broglia, Oscillations in Finite Quantum Systems,
               Cambridge Univ. Press, 2001.

\bibitem{IJMP-A-07} S.I. Bastrukov, H-K. Chang, \c S. Mi\c sicu, I.V. Molodtsova, D.V. Podgainy, Int. J. Mod. Phys.
            A 22 (2007) 3261.

\bibitem{HVDW-01} M.N. Harakeh, A. Van Der Woude, Giant Resonances: Fundamental High-Frequency Modes of Nuclear Excitation, Clarendon,  Oxford, 2001.


\bibitem{R-04} A. Richter, Nucl. Phys. A 731 (2004) 59.

\bibitem{Z-04} A. Zilges, Nucl. Phys. A 731 (2004) 249.

\bibitem{KPZ-06} U. Kneissl, N. Pietralla, A. Zilges,  J. Phys.  G 32 (2006) 217.


\bibitem{H-97}  R.D. Herzberg, et al, Phys. Lett. B 390 (1997) 49.

\bibitem{W-98} M. Wilhelm, et al, Phys. Rev. C 57 (1998) 577.

\bibitem{G-98} K. Govaert, et al,  1998 Phys. Rev. C 57 (1998) 2229.

\bibitem{H-99}  R.D. Herzberg, et al, Phys. Rev.  C 60 (1999) 051307.

\bibitem{B-00}  F. Bauwens, et al, Phys. Rev. C 62 (2000) 024302.

\bibitem{R-02}  N. Ryezayeva, et al. Phys. Rev. Lett. 89 (2002) 272502.

\bibitem{Z-02}  A. Zilges,  et al, Phys. Lett. B 542 (2002) 43.

\bibitem{H-03}  T. Hartmann, et al, Nucl. Phys. A 719 (2003) C308.

                T. Hartmann, et al, Phys. Rev. Lett. 85 (2004) 274; Phys. Rev. Lett. 93 (2004) 19.

\bibitem{E-05}  J. Enders, et al,  Acta Phys. Polon. B 36 (2005) 1077.

\bibitem{OZ-07} B. \"Ozel,  et al, Nuc. Phys. A 788 (2007) 385.


\bibitem{BM-90} S.I. Bastrukov, J.A. Maruhn,  Z. Phys. A 335 (1990) 139.

\bibitem{HE-77} G. Holzwarth, G. Eckart, Z. Phys. A 283 (1977) 219; Nucl. Phys. A 325 (1979) 1.

\bibitem{SH-78} H. Sagawa, G. Holzwarth, Prog. Theor. Phys. A 325 (1978) 1213.

\bibitem{JPB-78} J-P. Blaizot,  Phys. Lett. 78B (1978) 367.

\bibitem{NS-80} J.R. Nix and A.J.  Sierk, Phys. Rev.  C 21 (1980) 396; Phys. Rev.  C 25 (1982) 1068.

 \bibitem{HGWL-82} R. Hasse,  G. Ghosh, J. Winter, A. Lumbroso, Phys. Rev. C 25 (1981) 2771.

\bibitem{WA-81} C-Y. Wong,  N. Azziz, Phys. Rev. C 24 (1981) 2290.

\bibitem{MB-83} M. Brack, Phys. Lett. 123B (1983) 143.

\bibitem{DR-88} M. Di Toro, G. Russo,  Z. Phys. A 331 (1988) 381.

\bibitem{CHANDRA} S. Chandrasekhar, Hydrodynamic and Hydromagnetic Stability,  Clarendon, Oxford, 1961.

\bibitem{BMS-93}  S.I. Bastrukov, \c S. Mi\c sicu, A.V. Sushkov, Nucl. Phys. A 562 (1993) 191.

                  \c S. Mi\c sicu, J. Phys. G 14 (1995) 545.

                  \c S. Mi\c sicu, S.I. Bastrukov, Eur. Phys. J. A 13 (2002) 399.


\bibitem{SEM-81} S.F. Semenko, Sov. J. Nucl. Phys. 34 (1981) 356.

\bibitem{BM-88} E.B. Balbutsev, I.N. Mikhailov, J. Phys. G 14 (1988) 545.

                E.B. Balbutsev, I.V. Molodtsova, A.V. Unzhakova, Eur. Phys. Lett. 26 (1994) 499.


\bibitem{Lo} J. Kvasil, N. Lo Iudice,  Ch. Stoyanov, P. Alexa, J. Phys. G 29 (2003) 753.

\bibitem{SM-06} \c S. Mi\c sicu, Phys. Rev. C 73 (2006) 024301.

\bibitem{P-07} N. Paar, D. Vretenar, E. Khan, G. Colo, Rep. Prog. Phys. 70 (2007) 691.

\bibitem{MB-08} \c S. Mi\c sicu,  S.I. Bastrukov, in Fission and Properties of Neutron-Rich Nuclei, World Scientific,
                 Singapore 2008.


%%%%%%%%%%%%%%%%%%%%%%%%% LDM %%%%%%%%%%%%%%%%%%%%%%%%%%%%%%


\bibitem{MDB-71} R. Mohan, M. Danos, L.C. Biedeharn, Phys. Rev. C 3 (1971) 1740.


\bibitem{SIS-90} Y. Suzuki,  K. Ikeda, H. Sato, Prog. Theor. Phys. 83 (1990) 180.

%%%%%%%%%%%%%%%%%%%%% TORUS-M %%%%%%%%%%%%%%%%%%%%%%%%%%%%%%%%%%%%%%%%%

\bibitem{BMS-95}  S.I. Bastrukov, I.V. Molodtsova, V.M. Shilov, Phys. Rev. C 52 (1995) 1114.

\bibitem{BLM-97} S.I. Bastrukov,  J. Libert, I.V. Molodtsova, Int. J. Mod. Phys. E 6 (1997) 89.

\bibitem{RFH-91} S. Raman, L.W. Fagg, R. Hicks, in: Giant Magnetic Resonances, ed by Speth J, World Scientific, Singapore 1991.

%%%%%%%%%%%%%%%%%%%%%%%%%%%%%%%%%%%%%%%%%%%GR&&&&&&&&&&&&&&&&&&&&&&&&&&&

\bibitem{B-81} F.E. Bertrand, Nucl. Phys. A 354 (1981) 129.

\bibitem{TY-81} T. Yamagata, et al, Phys. Rev. C 23 (1981) 937.

H. L. Clark, D. H. Youngblood, Y.-W. Lui  Phys. Rev. C 54 (1996) 72.

\bibitem{LNCNR-96} C. Luttge, P. von Neumann-Cosel, F. Neumeyer, A. Richter,
 Nucl. Phys. A 606 (1996) 183.

%%%%%%%%%%%%%%%%%%%%%%%%% NS %%%%%%%%%%%%%%%%%%%%%%%%%%%%%%%%%%%%%

\bibitem{H-78} R. Hasse, Rep. Prog. Phys. 41 (1978) 1027.


\bibitem{MVH-88} P.N. McDermott, H.V. van Horn, C.D. Hansen, Astrophys. J.
 {\bf 325} (1988) 725.

\bibitem{BWP-99} S.I. Bastrukov, F. Weber, D.V. Podgainy, J. Phys. G 25 (1999) 107.

\bibitem{MPL-A} S.I. Bastrukov, H-K. Chang, G-T. Chen, I.V. Molodtsova, Mod. Phys. Lett. A 23 (2008) 477.





\end{thebibliography}
\end{document}